\PassOptionsToPackage{table,xcdraw}{xcolor}
\documentclass[sigconf,natbib=false,authorversion]{acmart}

\copyrightyear{2020}
\acmYear{2020}
\setcopyright{acmcopyright}\acmConference[WiSec '20]{13th ACM Conference on Security and Privacy in Wireless and Mobile Networks}{July 8--10, 2020}{Linz (Virtual Event), Austria}
\acmBooktitle{13th ACM Conference on Security and Privacy in Wireless and Mobile Networks (WiSec '20), July 8--10, 2020, Linz (Virtual Event), Austria}
\acmPrice{15.00}
\acmDOI{10.1145/3395351.3399420}
\acmISBN{978-1-4503-8006-5/20/07}
\acmSubmissionID{73}

\RequirePackage[abbreviate=true, dateabbrev=true, isbn=true, doi=true,
urldate=comp, url=true, maxcitenames=2, maxbibnames=9, backref=false, backend=biber,
style=ACM-Reference-Format, language=american]{biblatex}
\DeclareFieldFormat{titlecase}{#1}
\renewcommand{\bibfont}{\Small}

\usepackage[utf8]{inputenc}
\usepackage{lipsum}  
\usepackage{siunitx}
\usepackage[font=normalsize]{subcaption}

\usepackage{pgfplots}
\pgfplotsset{compat=newest}
\usetikzlibrary{plotmarks}
\usetikzlibrary{arrows.meta}
\usepgfplotslibrary{patchplots}
\DeclareMathAlphabet{\mathpzc}{OT1}{pzc}{m}{it}
\usepackage{grffile}
\usepackage{amsmath}
\pgfplotsset{plot coordinates/math parser=false}
\newlength\figureheight
\newlength\figurewidth

\newcommand{\MATLAB}{{\footnotesize MATLAB}}

\usepackage{acronym}
\makeatletter
\newcommand{\acreset}[1]{%
  \AC@reset{#1}%
}
\makeatother
\begin{document}

\acrodef{sdp}[SDP]{Secure Device Pairing}
\acrodef{iot}[IoT]{Internet of Things}
\acrodef{aic}[AIC]{Acoustic Integrity Code}
\acrodef{ook}[OOK]{on-off keying}
\acrodef{ask}[ASK]{amplitude-shift keying}
\acrodef{snr}[SNR]{signal-to-noise ratio}
\acrodef{ber}[BER]{bit error ratio}
\acrodef{mitm}[MITM]{machine-in-the-middle}
\acrodef{oob}[OOB]{out-of-band}
\acrodef{tls}[TLS]{Transport Layer Security}
\acrodef{los}[LOS]{line-of-sight}
\acrodef{sdk}[SDK]{software development kit}
\acrodef{art}[ART]{Android runtime}

\title{Acoustic Integrity Codes: Secure Device Pairing Using Short-Range
  Acoustic Communication}

\author{Florentin Putz}
\orcid{0000-0003-3122-7315}
\affiliation[obeypunctuation=true]{%
  \department[0]{Secure Mobile Networking Lab}\\
  \department[1]{Department of Computer Science}\\
  \institution{TU Darmstadt}, \country{Germany}
}
\email{fputz@seemoo.de}

\author{Flor Álvarez}
\orcid{0000-0002-0584-6406}
\affiliation[obeypunctuation=true]{%
  \department[0]{Secure Mobile Networking Lab}\\
  \department[1]{Department of Computer Science}\\
  \institution{TU Darmstadt}, \country{Germany}
}
\email{falvarez@seemoo.de}

\author{Jiska Classen}
\affiliation[obeypunctuation=true]{%
  \department[0]{Secure Mobile Networking Lab}\\
  \department[1]{Department of Computer Science}\\
  \institution{TU Darmstadt}, \country{Germany}
}
\email{jclassen@seemoo.de}

\renewcommand{\shortauthors}{F. Putz et al.}

\begin{abstract}
\ac{sdp} relies on an out-of-band channel to authenticate devices. This requires a
common hardware interface, which limits the use of existing \ac{sdp} systems.
We propose to use short-range acoustic communication for the initial pairing.
Audio hardware is commonly available on existing off-the-shelf devices and can be accessed from
user space without requiring firmware or hardware modifications.

We improve upon
previous approaches by designing \textit{\acp{aic}}: a modulation
scheme that provides message
authentication on the acoustic physical layer. We analyze their security and
demonstrate that we can defend against signal cancellation attacks by designing
signals with low autocorrelation. Our system can detect overshadowing attacks
using a ternary decision function with a threshold.
In our evaluation of this \ac{sdp} scheme's security and robustness, we achieve a bit error ratio
below 0.1\% for a net bit rate of \SI{100}{bps} with
a  \ac{snr} of \SI{14}{dB}. Using our open-source proof-of-concept
implementation on Android smartphones, we demonstrate pairing between different
smartphone models.
\end{abstract}



\maketitle

\newpage
\section{Introduction}

An increasing
number of ubiquitous computing devices require secure provisioning and pairing mechanisms. During the setup of
cyber-physical systems, consumers still struggle with constructing secure
communication channels, as those require preexisting \textit{security
contexts}, e.g. shared public keys when using asymmetric cryptography.
Establishing such a prior
security context is a critical step to ensure the communication's security.

One way of establishing a new security context is \ac{sdp}. The devices pair in an ad-hoc manner and establish an authenticated key.
In contrast to public key infrastructures (e.g., x.509 or OpenPGP), \ac{sdp} does not
require any \textit{trusted third parties}, where Alice has to trust other
entities that help her initialize a security context with Bob.
Therefore, \ac{sdp} is well-suited for
offline and private scenarios, emergency settings, or bootstrapping new deployments that are not part
of any public key infrastructure. There is growing research in using \textit{physical device proximity} to support
\ac{sdp}, by either using a location-limited communication channel or extracting keys from
measuring the environment~\autocite{Hu2018}.

\ac{sdp} requires a common hardware interface that both devices use for pairing,
which limits the applicability of existing \ac{sdp} schemes.
Many
\ac{sdp} schemes use 
displays, cameras, vibration motors,
accelerometers, infrared transducers, or wireless near-field
communication~\autocite{Fomichev2018,Fomichev2019,Hu2018}.
A commonly available hardware interface that is rarely used for \ac{sdp} in
practice is audio
via speakers and microphones. \textit{Acoustic communication} requires only minimal user interaction, which
increases usability and reduces potential failure points. In contrast to
electromagnetic wireless communication such as Wi-Fi or Bluetooth, acoustic
communication requires no complex network configuration and can be implemented
as user space software, even with physical-layer capabilities. This allows us to perform acoustic communication on existing
off-the-shelf devices without hardware modification, reducing deployment costs~\cite{Lopes2001}.

Previous approaches that used audio for \ac{sdp}, such as
``HAPADEP''~\autocite{Soriente2008}, require a manual 
verification phase for security reasons, which is
error-prone and reduces usability.
We design a secure acoustic communication protocol that requires less
security-critical user interaction by incorporating
the recent research direction of \textit{physical-layer security}. These
techniques consider security at the lowest layer using the physical properties
of the wireless radio channel. Specifically, we use \textit{Integrity Codes},
which were proposed by \citeauthor{Capkun2008} to provide message integrity in
the presence of active attackers on the radio channel~\autocite{Capkun2008}.
Using Integrity Codes, we 
eliminate the need for a separate verification step, which speeds up the pairing
process and increases usability.
Integrity Codes can be vulnerable to
\textit{signal cancellation} attacks~\autocite{Ghose2018,Hou2015,Pan2017}. We
therefore analyze this threat and propose countermeasures. Our resulting design
improves Integrity Codes by mitigating signal cancellation attacks.

\acreset{aic}
Our main contribution is the design, implementation and evaluation of \acp{aic},
which we use for \ac{sdp}. This work combines the independent  
research fields of \ac{sdp}, acoustic communication, and physical-layer security. To
the best of our knowledge, Integrity Codes have not been applied to acoustic
communication before.
Our individual contributions are:

\begin{itemize}
  \item \textbf{Analysis of Signal Cancellation Attacks.} We show that signal
    cancellation attacks fail for signals with low
    autocorrelation. We propose system parameters that improve Integrity Codes.
  \item \textbf{Design of Acoustic Integrity Codes.} We use Integrity Codes to
    secure acoustic communication.
  \item \textbf{Evaluation.} We evaluate the security and robustness of AICs
    using simulations.
  \item \textbf{Design of an \ac{sdp} scheme using AICs.} We apply AICs to
    design an acoustic SDP scheme.
  \item \textbf{Implementation of a prototype for modern Android devices.} We
    implement an open-source proof-of-concept for Android smartphones.
\end{itemize}

Our work is structured as follows: After introducing related work and Integrity Codes
in \autoref{sec:related-work}, we present our design in \autoref{sec:design} and
our implementation in \autoref{sec:implementation}. Then, we analyze the
security of \acp{aic} in \autoref{sec:security-analysis} and evaluate
them in \autoref{sec:evaluation}. Finally, we conclude our work in \autoref{sec:conclusion}.

\section{Related Work}\label{sec:related-work}
In this section, we describe \ac{sdp} and present related
work using acoustic communication to perform \ac{sdp}. We also introduce
Integrity Codes.

\subsection{Secure Device Pairing}

\ac{sdp} enables multiple devices
with no prior security context to establish a secure communication channel over
an untrusted channel. We only consider two devices $A$ and $B$ belonging to Alice and Bob,
respectively. The devices want to communicate over an a priori insecure communication
channel. They use an \textit{\ac{oob} channel} to authenticate
a key exchange, which they can then use to construct a secure communication
channel using a standard cryptographic protocol such as \ac{tls}.

The audio channel
can be used as a \textit{location-limited channel} to perform this key
exchange~\autocite{Balfanz2002}. 
Goodrich et al.~\autocite{Goodrich2009,Goodrich2006} developed an \ac{sdp} system called
``Loud\&Clear'', which requires the user to detect whether two
computer-generated speech sequences are identical. Soriente et
al.~\autocite{Soriente2008} presented ``HAPADEP'', in which the devices encode
their public keys as short audible melodies that the other device can decode.
They use a second verification phase to detect \ac{mitm} attacks, which requires
active user participation.
Halperin et al.~\autocite{Halperin2008} developed an \ac{sdp} system called
``Zero-Power Sensible Key Exchange'' involving acoustic communication for
implantable medical devices (IMDs). The IMD generates a symmetric
session key and transmits it as an audible sound wave to the external device via
a piezo element. This OOB channel has to be secret, since their system lacks
eavesdropping protection. Halevi and
Saxena~\autocite{Halevi2010} showed that eavesdropping is
possible even with off-the-shelf equipment, using digital signal processing.
Claycomb and Shin~\autocite{Claycomb2009} devised an acoustic \ac{sdp} method
called ``UbiSound'', which uses a single unidirectional audio transmission. The
user is responsible for aborting the pairing process in case of malicious interference.
Mayrhofer et al.~\autocite{Mayrhofer2013} presented ``UACAP'' as a general \ac{sdp} implementation
that is designed to support multiple OOB channels such as 2D barcodes, manual
string comparison, and audio (based on ``HAPADEP'').
Han et al.~\autocite{Han2014} proposed the \ac{sdp} protocol ``MVSec'', using
either an audio or a visual channel as the OOB channel to pair smartphones with
cars. This pairing protocol happens mainly over an in-band Bluetooth channel.
The audio channel is used as the OOB channel to bidirectionally transfer
truncated commitments to the public keys.

In contrast to our design, these \ac{sdp} schemes usually require additional user
interaction to defend against active attackers.
To the best of our knowledge,
there is no publicly available acoustic \ac{sdp} implementation for current iOS or Android devices.

Apart from the audible sound spectrum, the inaudible ultrasound spectrum has
also been used as part of the secure pairing
process~\autocite{Kindberg2003,Mayrhofer2006,Mayrhofer2007}.
The ultrasound spectrum, however, is
not suitable to secure commercial off-the-shelf devices, because this requires
additional hardware.
Apart from using
acoustic communication directly, previous research on \ac{sdp} also utilized the
audio channel for demonstrative identification~\autocite{Peng2009}, as part of
an audiovisual pairing scheme~\autocite{Prasad2008}, or for ambient
sensing~\autocite{Miettinen2014,Quach2014,Schuermann2013}. 

\subsection{Integrity Codes}
Over the last decade, the research community has investigated whether security goals
such as \textit{authentication} and \textit{integrity} protection can be
realized on the physical layer. As these physical-layer security techniques
usually do not assume a prior security context, they are well-suited to protect
\ac{sdp}. In this section, we introduce the \textit{Integrity Code} physical-layer
security primitive~\autocite{Capkun2008}, which we use to secure
acoustic communication.

\begin{figure}[!b]
  \centering
  \includegraphics[page=2,width=\columnwidth]{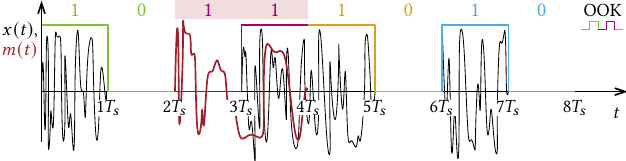}
  \Description{An on-off-keyed signal consisting of noise in the on slots and
  silence in the off slots. An attacker interferes with a second signal, which
  adds to the silence in one of the off slots and flips it to an on slot. The
  receiver can detect this, since this violates the Manchester Code.}
  \caption[I-Code with an Attacker]{Integrity Code signal representing
    the data \texttt{1011}. An attacker interferes using the red signal.
    This flips the third bit from $0 \to 1$, which can be detected.}
  \label{fig:icode-with-attacker}
\end{figure}

\citeauthor{Capkun2008} established the concept of \textit{Integrity Codes
  ("I-Codes")}~\cite{Capkun2008}, which is a modulation scheme that protects the
message integrity on the wireless
physical layer without requiring any
shared key material. Instead, integrity and authentication can be protected when
the receiver knows that:
\begin{enumerate}
\item the sender is currently transmitting and
\item the sender is in the receiver's range.
\end{enumerate}
\autoref{fig:icode-with-attacker} illustrates Integrity Codes. The transmitter first applies a unidirectional error code (e.g., Manchester Code
or Berger Code), which can detect one direction of bit flips (from $0 \to 1$). Then,
the transmitter performs \ac{ook} using the encoded data.
\ac{ook} is a form of \ac{ask}, where a $0$ is
represented by the absence of a carrier wave and a $1$ is represented by the
presence of a carrier wave. Instead of using a deterministic carrier signal,
however, we use a stochastic signal in each on slot. The idea is that random
signals cannot be 
cancelled by an active attacker via destructive interference. An attacker is not able to change
any bit from $1 \to 0$. Any other modification of the message can be detected at
the receiver using unidirectional error codes.
Integrity Codes have been used to assist in \ac{sdp} with radio
communication~\autocite{Gollakota2011,Shen2016}. We use Integrity Codes to secure acoustic communication.

\section{Design}\label{sec:design}

In this section, we present the design of our \ac{sdp} scheme, which uses short-range acoustic communication.
We design \acp{aic} to secure this communication on the
physical layer.
Our main goal is to securely transmit public key material $d$ (or shorter
commitments such as hash values) from Alice's
device $A$ to Bob's device $B$, even in the presence of an active adversary
Mallory, who tries to manipulate this 
communication using her devices $M_n$. We focus on unidirectional \ac{sdp} for the private and social
application classes and consider pairing in
the other direction as an optional subsequent but separate step that works in
the same way. The public key material can then be used to initialize a security
context between the devices.

Specifically, our system shall achieve \textit{message
  authentication}
of the transmitted public
key material $d$, which consists of the following two security properties
\cite[25]{Katz1996}:
\begin{itemize}
  \item \textbf{Identification of the sender:} $B$ is able to verify whether the
    message $d$ originated from $A$.
  \item \textbf{Integrity:} $B$ can detect whether the message $d$ was modified
    during transmission. This is implied by the first property, since then $A$
    would no longer be the message's originator.
\end{itemize}
Confidentiality or availability protection is out of scope.

\begin{figure}[!b]
  \centering
  \includegraphics[page=5,width=\columnwidth]{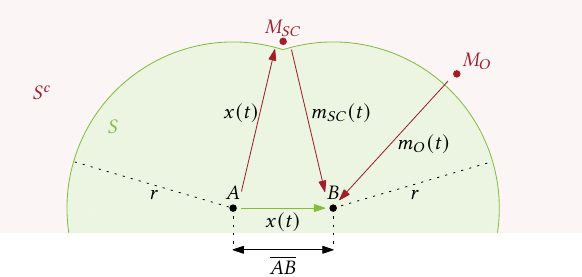}
  \Description{Two devices (A and B) in a 2D plane. The green safe area is shown as
    the union of two circles with radius r centered at each of the device
    locations. Outside is the red insecure area. Two malicious devices are
    located in the insecure area. Arrows show the signal propagation of
    transmissions from A to B, from A to the signal cancellation attacker, from
    the signal cancellation attacker to B, and from the overshadowing attacker
    to B.}
  \caption[System Model]{This system model shows a top-down perspective of the
    devices $A$ and $B$ with the safe area $S$ around them (green).
    Possible adversaries can perform signal cancellation attacks with device $M_{SC}$
    (see \autoref{sec:signal-cancellation}) or overshadowing attacks with device $M_O$ (see \autoref{sec:overshadowing})
    in the insecure area $S^c$ (red)
    outside the secure area $S$.} 
  \label{fig:system-model}
\end{figure}

We use an \textit{acoustic channel} as the physical channel, i.e., we transmit
information by modulating a mechanical pressure wave generated by $A$ using a
speaker~\autocite{Fomichev2018}.
The receiver $B$ records this using a microphone.
\autoref{fig:system-model} shows the devices $A$ and $B$ and their
environment during the pairing process.
We require that Alice and Bob perform the pairing
while being in proximity, such as by standing next to each other.
We denote the distance between the devices $A$ and $B$ as $\overline{AB}$.
We denote the immediate area around Alice and Bob as $S$, the \textit{safe
  area}. The remaining \textit{unsafe area} is $S^c$.
We model $S$ using two spheres centered at $A$ and $B$,
respectively, with radius $r$. 
\subsection{Assumptions}

We assume that Alice and Bob control the safe area, i.e., there are no
malicious devices in $S$.
We argue that this assumption is realistic for the
private application class such as when pairing devices at home. For
the social application class, this means that pairing should not be
performed in crowded areas, where an attacker could be close.
For the model parameters, we assume $r > \SI{40}{cm} > \overline{AB}$. These
parameters can be adjusted depending on the specific use case. We study
the security implications in \autoref{sec:security-analysis}.
We assume that device $A$ is equipped with a speaker and that device $B$ is
equipped with a microphone. 
Microphones and speakers
are commonly available on commercial off-the-shelf devices.
We assume that the devices $A$ and $B$ are not compromised, i.e., the software and
hardware performing the \ac{sdp} scheme are not controlled by Mallory.
Finally, we assume that $A$ and $B$ know the public protocol parameters, which
we explain in the next sections.

\subsection{Adversary Model}\label{sec:adversary-model}
Making realistic assumptions on the attacker's capabilities is crucial to
effectively design a secure system~\autocite{Ferguson2010}. Whereas weak attacker models can underestimate the
threats, a very strong attacker model can lead to an overly complicated system
design or require more advanced hardware, which could hinder
adoption~\autocite{Poepper2011}. We make practical and realistic
assumptions. 

Research on wireless network security often uses a \emph{Dolev-Yao attacker
  model}~\autocite{Dolev1983}, which assumes a very strong attacker,
who is able to 
control and manipulate all messages on the network. 
It is, however, not necessarily realistic to assume that the attacker can \textit{freely}
modify or annihilate the wireless signals at the receiver's
antenna~\autocite{Poepper2011}.
We therefore use a weaker but more realistic Dolev-Yao attacker model, which
is typically used when applying Integrity Codes~\autocite{Capkun2008,Gollakota2011,Shen2016}.

The main goal of the attacker (Mallory) is to impersonate Alice. Mallory
wants that Bob accepts her key instead of Alice's key.
Mallory can
\textit{eavesdrop} all signals (passive). She can also \textit{send her own
  signals} (active),
which superimpose with the legitimate signal at the receiver's antenna.
Mallory's signal transmissions are still bound by the same physical signal
propagation laws that also govern legitimate transmissions, i.e., they arrive at
the receiver after a propagation delay with a phase shift.
We assume that Mallory can
only operate outside the secure area $S$, which is controlled and observed by
Alice and Bob via proximity. We assume that she cannot
trivially disable the communication channel by shielding $A$'s signals from 
propagating to $B$ with a physical barrier. 

Mallory may control any number $N$ of devices $\big\{M_n: n \in \left\{1, \dots,
  N\right\}\big\}$,
which
are placed anywhere outside the secure area $S$.
We denote as $x(t)$ the signal that $A$ transmits to $B$. We denote as $m_n(t)$ the signal
that $M_n$ transmits.
These signals are affected by
the acoustic channel $H$, which attenuates and delays the signal.
We also consider additional noise $v(t)$. Then, $B$
receives the following superposition of all these signals:
\begin{align}
  y(t) = \underbrace{H_{A \to B}\left\{ x(t) \right\}}_{\text{legitimate signal }x'(t)} + \sum_{n=1}^{N}   \underbrace{H_{M_n \to B}\left\{ m_n(t) \right\}}_{\text{attacker } m'_n(t)} + \underbrace{v(t)}_{\text{noise}}\label{eq:received-signal} 
\end{align}
When designing our communication system, we account for signal
cancellation, bit flipping, and overshadowing attacks.
We analyze these types of attacks in \autoref{sec:security-analysis}.

\subsection{Secure Device Pairing Scheme}\label{sec:design-pairing-scheme}

Alice uses her device $A$ to
transmit some public key material $d$ to Bob's device $B$. The pairing process
consists of the following steps:

\begin{enumerate}
  \item Alice initializes the pairing process on her device $A$. The device
    now repeatedly broadcasts $d$ on the acoustic channel using \acp{aic}.
  \item Alice tells Bob that he can start receiving data now.
  \item Bob accepts the pairing process on his device $B$.
  \item Device $B$ receives the key material over the acoustic channel.
  \item Device $B$ notifies Bob that it received the key. The transmission was
    either successful or there was an error due to background noise or an
    attacker.
  \item Bob tells Alice that he finished the pairing process.
  \item Alice stops the transmission on her device $A$.
\end{enumerate}

The \ac{sdp} process is successful if there was no transmission error. If the environmental
noise is too high or if there is an attacker, the transmission fails and
they can try again at another location. We design \acp{aic} to provide message
authentication of the communication on the physical layer.

\acp{aic} require that the receiver is aware of an ongoing transmission.
We could use additional signaling on the physical layer to automate the manual steps (2),
(3), (6) and (7), but this signaling could be modified by the attacker.
We cannot secure this signaling, since we assume that we have no prior security
context.
We also cannot use \acp{aic} to secure this signaling, since \acp{aic} require that the
receiver always knows that the legitimate transmitter is active.

\begin{figure*}[p]
  \centering
  \makebox[\textwidth][c]{
    \includegraphics[scale=0.9,page=1]{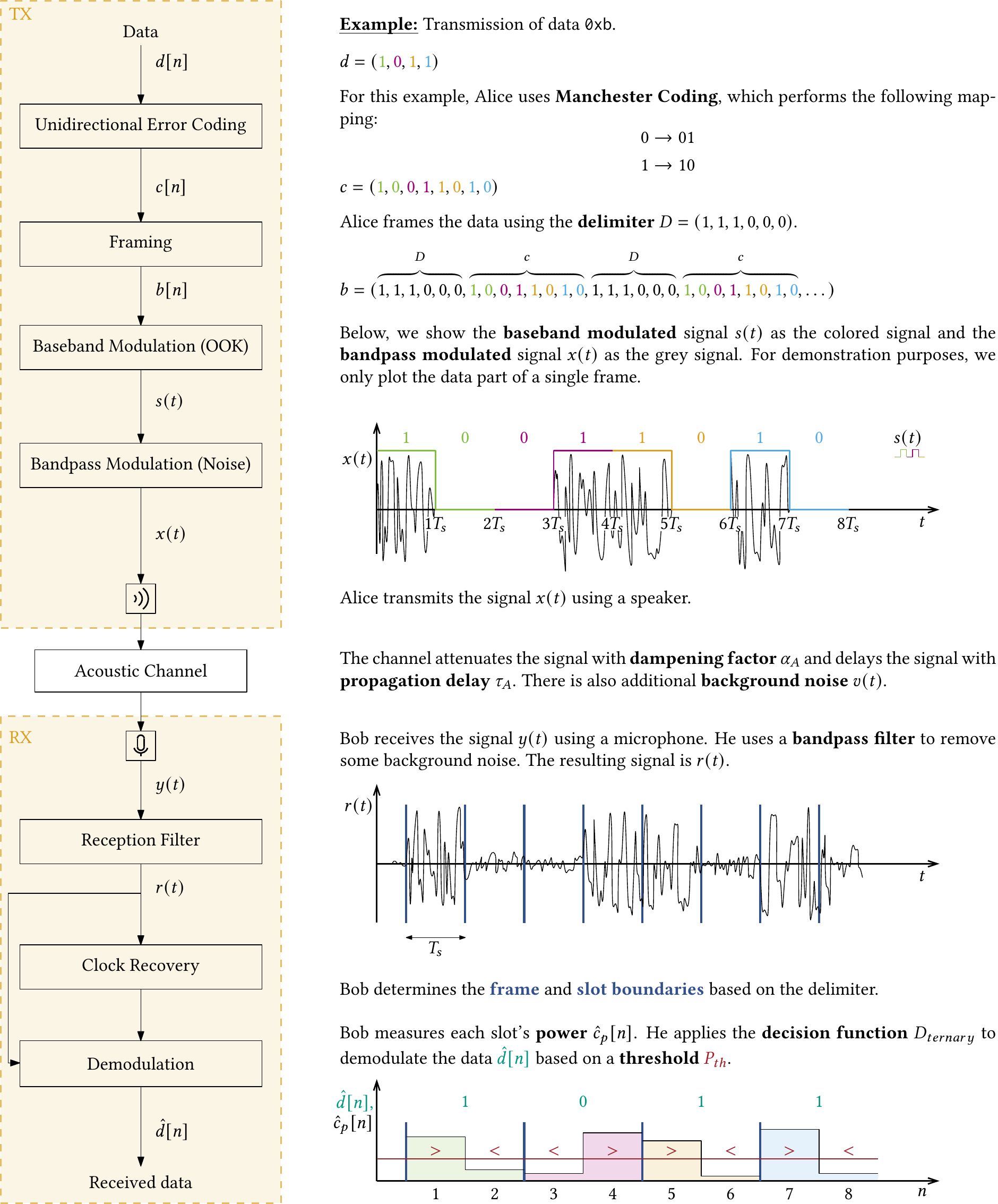}
  }
  \Description{A vertical block diagram is shown to the left, spanning the whole
  page height. The right side shows an example transmission and reception of the
  data 1 0 1 1, where each step is aligned with the corresponding step in the
  block diagram. This diagram visualizes the description from \autoref{sec:aic}.}
  \caption[Overview and Example of \ac{aic} Signals]{Construction, transmission,
    reception, and decoding of \ac{aic} signals, including a complete example.}
  \label{fig:comm-example}
\end{figure*}

\subsection{Acoustic Integrity Codes}\label{sec:aic}

We now explain how we secure acoustic communication using the \ac{aic}
modulation scheme.
This is the foundation of our \ac{sdp} scheme.
\acp{aic} apply the concept of Integrity Codes~\autocite{Capkun2008} to acoustic
signals.
We use Integrity Codes to defend against signal cancellation and overshadowing 
attacks.

\autoref{fig:comm-example} illustrates transmission and reception of \acp{aic}.
The transmitter $A$ encodes the data using unidirectional error coding (e.g.,
Manchester Coding) and frames the data by prepending a delimiter header 
$D[n]=(1,1,1,0,0,0)$.
\citeauthor{Capkun2008} have shown that this delimiter is
optimal~\cite{Capkun2008}.
After framing, $A$ converts the time discrete sequence $b[n]$ (consisting of
repeated frames) into a time continuous baseband signal $s(t)$ using baseband
\ac{ook}.
As a result, each bit $b[n]$ corresponds to a time slot of
duration $T_s$ during which $s(t)$ has a constant value of either $0$ or $1$.
The signal's gross bit rate,
including coding overhead and the delimiter, is $R_g = 1/T_s$.
The net bit rate $R_n \approx R_g/2$ describes the
effective number of bits that can be transmitted per second. 

Finally, $A$ generates the bandpass signal $x(t)$ in the frequency band
$[f_{\text{low}},f_{\text{high}}]$.
Instead of using a deterministic carrier signal, $A$ modulates a
\textit{stochastic ``carrier'' signal} $w(t)$, sampled from a random process 
$\{W(t)\}$.
Unless otherwise noted, we use a white Gaussian noise process
$\{W_{\text{WGN}}(t)\}$. 
This step differs from conventional modulation schemes such as \ac{ask}, where $s(t)$ is
used to modulate a deterministic carrier signal. This randomness is essential
for integrity protection, as stochastic signals with low autocorrelation cannot
be cancelled out by an attacker (see \autoref{sec:security-analysis}).

After modulation, $A$ transmits the signal $x(t)$ through the acoustic 
channel $H_{A \to B}$ using a speaker:
\begin{align}
  x'(t) = H_{A \to B}\left\{ x(t) \right\} = \alpha_A x(t - \tau_A)\label{eq:x-prime}
\end{align}
In our system model, we expect to have a
strong \ac{los} component due to the devices' proximity.
Our approximate channel model
accounts for the attenuation $\alpha_A$ and propagation delay $\tau_A$ on the
\ac{los} path.
The propagation delay is proportional to
the distance between the devices and satisfies the relation $\tau_A = \overline{AB} / c_s$,
where $c_s$ is the speed of sound in our transmission medium.
The transmission is also subject to additive noise $v(t)$ resulting from
sound sources in the environment and from thermal noise in the
electrical components.

$B$ records and receives the resulting signal $y(t)$, as shown in \autoref{eq:received-signal}.
After filtering out background noise using a bandpass filter, $B$ performs synchronization to
recover the frame boundaries based on their delimiter and then demodulates this
signal using a decision function. In conventional modulation schemes such as \ac{ask},
we would not consider an attacker at the physical layer and therefore always
decide on one of two possible states $S_{\text{binary}}=\left\{0,1\right\}$, e.g., based on maximum
likelihood. Such a decision function $D_{\text{binary}}$, however, is vulnerable to an
overshadowing attack, which we analyze in \autoref{sec:overshadowing}.
Instead, our decision function $D_{\text{ternary}}$ considers an attacker at the physical layer and
decides on one of three possible states $S_{\text{ternary}} = \left\{0,1,\varepsilon  \right\}$,
where $\varepsilon$ signals an error:
\begin{align}
  &D_{\text{ternary}}(p_1, p_2) =
  \begin{cases}
    0, & \text{if } p_1 < P_{\text{th}} \text{ and } p_2 > P_{\text{th}}\\
    1, & \text{if } p_1 > P_{\text{th}} \text{ and } p_2 < P_{\text{th}}\\
    \epsilon, & \text{otherwise.}
  \end{cases} \label{eq:d-ternary}
\end{align}
This decision function compares both slot
powers of a Manchester pair with a threshold $P_{\text{th}}$, which influences both the
robustness and security of \acp{aic}.
We typically do not work with the absolute detection threshold, but with
the detection threshold relative to the noise floor $\text{SNR}_{\text{th}}$.
Security-wise, it should be as low as possible.

\section{Implementation}\label{sec:implementation}
\acreset{aic}
\acreset{sdp}
In this section, we show how \acp{aic} and our resulting \ac{sdp} scheme can be
implemented.
We developed our system for two different platforms:
\begin{enumerate}
\item \textbf{An implementation in \MATLAB{}} for simulation and evaluation. 
\item \textbf{A proof-of-concept on Android smartphones} for practical experiments, using the
  Kotlin programming language. 
\end{enumerate}

\begin{figure}[!b]
  \centering
  \includegraphics[page=4,scale=0.8]{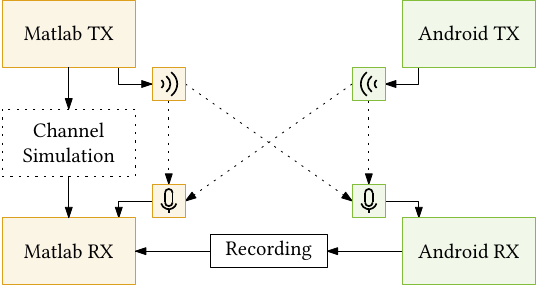}
  \Description{Block diagram showing the Matlab transmitter, the Matlab
    receiver, the Android transmitter, and the Android receiver. Arrows indicate
  possible transmissions between these implementations. The channel can be
  simulated from the Matlab transmitter to the receiver. It is also possible to
  play a signal from the Matlab transmitter and receive it using either the
  Matlab receiver or the Android receiver. It is also possible to play a signal
  from the Android transmitter and receive it using the Matlab receiver or the
  Android receiver. Lastly, it is possible to save a recording from the Android
  receiver and load it in the Matlab receiver.}
  \caption[Implementation Overview]{Overview of \MATLAB{}
    and Android implementations.
  }
  \label{fig:implementations-overview}
\end{figure}

\autoref{fig:implementations-overview} shows an overview of both
implementations. We can transmit and receive \acp{aic} using either implementation.
It is possible
to record the signals in the Android prototype into a WAV file and analyze this
using \MATLAB{}.

\subsection{Simulation}\label{sec:simulation-matlab}

We use \MATLAB{} version 9.4 R2018a to simulate \acp{aic}.
Our implementation can
generate, transmit, receive, and demodulate \ac{aic} signals using the computer's
speaker and microphone.
Alternatively, we can also simulate a transmission using
an \textit{additive white Gaussian noise (AWGN)} channel. We implement all steps
shown in \autoref{fig:comm-example}. 

\begin{figure}[!b]
  \centering
  \setlength\figureheight{2.6cm}
  \setlength\figurewidth{0.68\columnwidth}
%
%
\begin{tikzpicture}

\begin{axis}[%
width=0.967\figurewidth,
height=\figureheight,
at={(0\figurewidth,0\figureheight)},
scale only axis,
point meta min=-156.451851,
point meta max=-78.697384,
axis on top,
separate axis lines,
every outer x axis line/.append style={black},
every x tick label/.append style={font=\color{black}},
every x tick/.append style={black},
xmin=0.019444,
xmax=3.983435,
xtick={0.500000,1.000000,1.500000,2.000000,2.500000,3.000000,3.500000},
xlabel={Time in s},
every outer y axis line/.append style={black},
every y tick label/.append style={font=\color{black}},
every y tick/.append style={black},
ymin=10.000000,
ymax=22.000000,
ytick={10.000000,12.000000,14.000000,16.000000,18.000000,20.000000,22.000000},
ylabel={Frequency in kHz},
axis background/.style={fill=white},
clip mode=individual,tick label style={font=\tiny},label style={font=\small},colorbar style={font=\small}, title style={font=\small\bfseries},,
colormap={mymap}{[1pt] rgb(0pt)=(0.2422,0.1504,0.6603); rgb(1pt)=(0.25039,0.164995,0.707614); rgb(2pt)=(0.257771,0.181781,0.751138); rgb(3pt)=(0.264729,0.197757,0.795214); rgb(4pt)=(0.270648,0.214676,0.836371); rgb(5pt)=(0.275114,0.234238,0.870986); rgb(6pt)=(0.2783,0.255871,0.899071); rgb(7pt)=(0.280333,0.278233,0.9221); rgb(8pt)=(0.281338,0.300595,0.941376); rgb(9pt)=(0.281014,0.322757,0.957886); rgb(10pt)=(0.279467,0.344671,0.971676); rgb(11pt)=(0.275971,0.366681,0.982905); rgb(12pt)=(0.269914,0.3892,0.9906); rgb(13pt)=(0.260243,0.412329,0.995157); rgb(14pt)=(0.244033,0.435833,0.998833); rgb(15pt)=(0.220643,0.460257,0.997286); rgb(16pt)=(0.196333,0.484719,0.989152); rgb(17pt)=(0.183405,0.507371,0.979795); rgb(18pt)=(0.178643,0.528857,0.968157); rgb(19pt)=(0.176438,0.549905,0.952019); rgb(20pt)=(0.168743,0.570262,0.935871); rgb(21pt)=(0.154,0.5902,0.9218); rgb(22pt)=(0.146029,0.609119,0.907857); rgb(23pt)=(0.138024,0.627629,0.89729); rgb(24pt)=(0.124814,0.645929,0.888343); rgb(25pt)=(0.111252,0.6635,0.876314); rgb(26pt)=(0.0952095,0.679829,0.859781); rgb(27pt)=(0.0688714,0.694771,0.839357); rgb(28pt)=(0.0296667,0.708167,0.816333); rgb(29pt)=(0.00357143,0.720267,0.7917); rgb(30pt)=(0.00665714,0.731214,0.766014); rgb(31pt)=(0.0433286,0.741095,0.73941); rgb(32pt)=(0.0963952,0.75,0.712038); rgb(33pt)=(0.140771,0.7584,0.684157); rgb(34pt)=(0.1717,0.766962,0.655443); rgb(35pt)=(0.193767,0.775767,0.6251); rgb(36pt)=(0.216086,0.7843,0.5923); rgb(37pt)=(0.246957,0.791795,0.556743); rgb(38pt)=(0.290614,0.79729,0.518829); rgb(39pt)=(0.340643,0.8008,0.478857); rgb(40pt)=(0.3909,0.802871,0.435448); rgb(41pt)=(0.445629,0.802419,0.390919); rgb(42pt)=(0.5044,0.7993,0.348); rgb(43pt)=(0.561562,0.794233,0.304481); rgb(44pt)=(0.617395,0.787619,0.261238); rgb(45pt)=(0.671986,0.779271,0.2227); rgb(46pt)=(0.7242,0.769843,0.191029); rgb(47pt)=(0.773833,0.759805,0.16461); rgb(48pt)=(0.820314,0.749814,0.153529); rgb(49pt)=(0.863433,0.7406,0.159633); rgb(50pt)=(0.903543,0.733029,0.177414); rgb(51pt)=(0.939257,0.728786,0.209957); rgb(52pt)=(0.972757,0.729771,0.239443); rgb(53pt)=(0.995648,0.743371,0.237148); rgb(54pt)=(0.996986,0.765857,0.219943); rgb(55pt)=(0.995205,0.789252,0.202762); rgb(56pt)=(0.9892,0.813567,0.188533); rgb(57pt)=(0.978629,0.838629,0.176557); rgb(58pt)=(0.967648,0.8639,0.16429); rgb(59pt)=(0.96101,0.889019,0.153676); rgb(60pt)=(0.959671,0.913457,0.142257); rgb(61pt)=(0.962795,0.937338,0.12651); rgb(62pt)=(0.969114,0.960629,0.106362); rgb(63pt)=(0.9769,0.9839,0.0805)},
colorbar,
colorbar style={ylabel style={font=\small\color{black}}, ylabel={Power in dBFS}, yticklabel style={font=\small}}
]
\addplot [forget plot] graphics [xmin=0.019444, xmax=3.983435, ymin=-0.003125, ymax=22.053125] {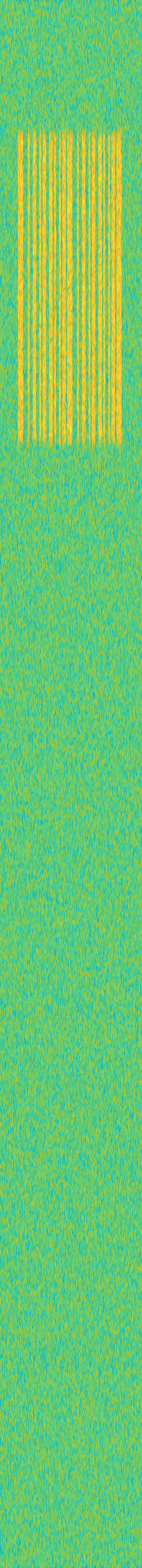};
\end{axis}
\end{tikzpicture}%
  \Description{Spectrogram plot showing time on the horizontal axis and
    frequency on the vertical axis. The power of the signal is encoded as the
    pixel color in the plot. The plot shows multiple rectangles containing
    high power. These rectangles all have the same height corresponding to the
    frequency band 16 kHz to 20 kHz. The width of each rectangle varies: Long
    rectangles correspond to the delimiter, medium and small rectangles
    correspond to regular on slots.}
  \caption[Spectrogram of an \ac{aic} Signal]{Spectrogram of the \ac{aic} signal $y(t)$. The frequency
    resolution is \SI{50}{Hz}.}
  \label{fig:simulation-spectrogram}
\end{figure}

\autoref{fig:simulation-spectrogram} shows the spectrogram of the signal $y(t)$,
which visualizes the power per frequency
over time.
In this example, the \ac{aic} signals' energy is concentrated in the frequency band 
$[\SI{16}{kHz},\SI{20}{kHz}]$ with a SNR of \SI{14}{dB}.
We can clearly identify the three delimiters (the wider rectangles at
the beginning, middle, and end), and the on and off slots in between.

\subsection{Proof of Concept}\label{sec:prototype-android}
Our proof-of-concept implementation runs on Android devices and is available as
open-source software~\cite{AICPrototype}.
We can receive and
transmit \ac{aic} signals, or we can record a WAV file for later
analysis. We use the Android \ac{sdk}
version 28 and the Kotlin programming language version 1.3~\cite{Kotlin2019}
to write Android 
applications.
We generate Java 8 compatible
bytecode, which the Android SDK 
translates to Dalvik bytecode for use on the \ac{art} on Android
devices. Our Android application requires a minimum Android API level of 21,
meaning that it supports all devices with Android 5.0
(released in 2014)
or higher. This is not a limitation of our design, but allows for easier
development of a prototype by
accessing more API features.

\begin{figure}[!b]
  \begin{subfigure}{0.49\columnwidth}
    \centering
    \includegraphics[width=0.93\columnwidth]{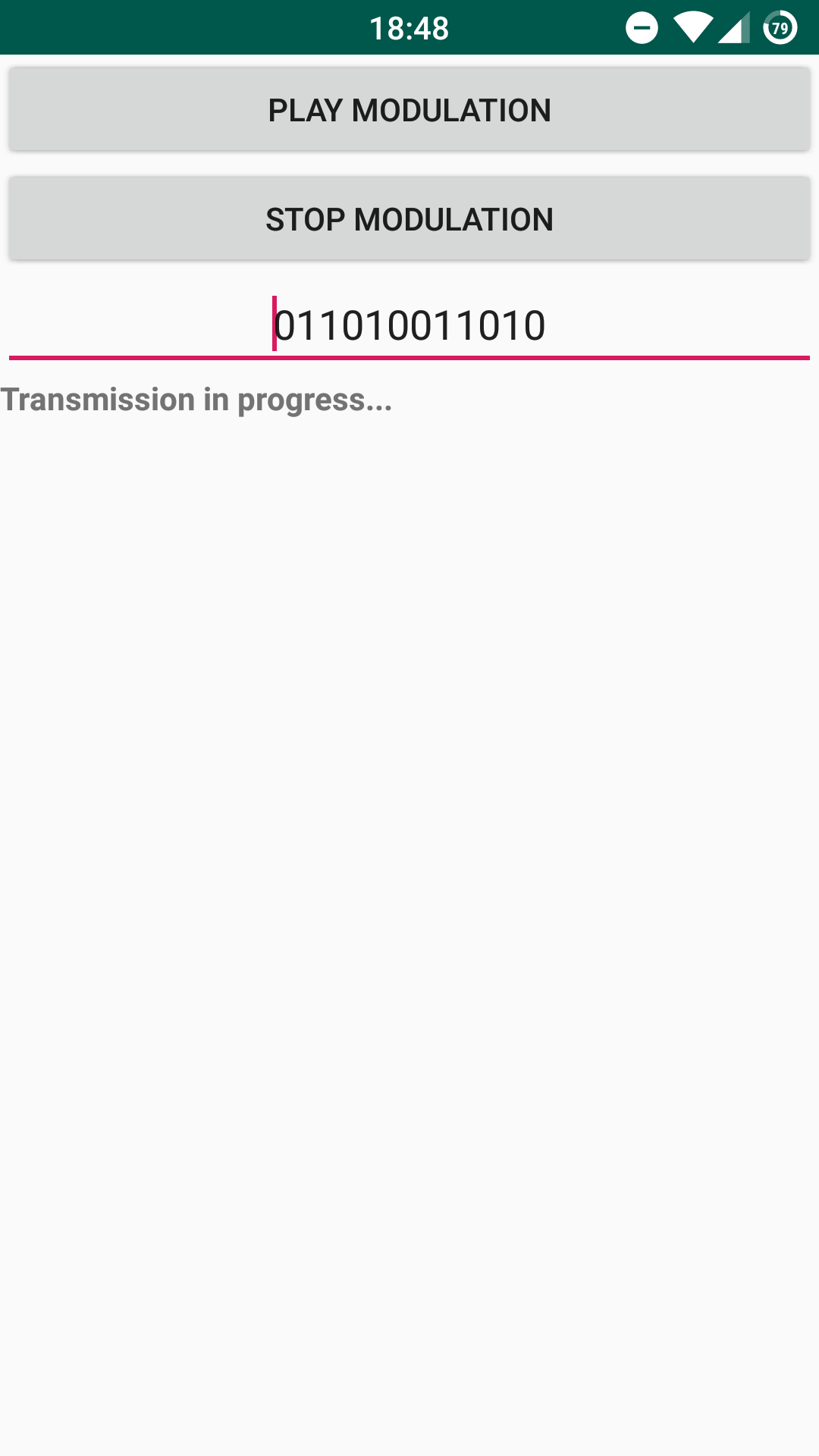}
    \Description{Android screenshot showing two buttons labelled ``Play
      modulation'' and ``Stop modulation'', a text input field containing
      ``011010011010'', and a label with text ``Transmission in progress...''.}
    \caption{Transmission}
    \label{fig:android-txfragment}
  \end{subfigure}
  \begin{subfigure}{0.49\columnwidth}
    \centering
    \includegraphics[width=0.93\columnwidth]{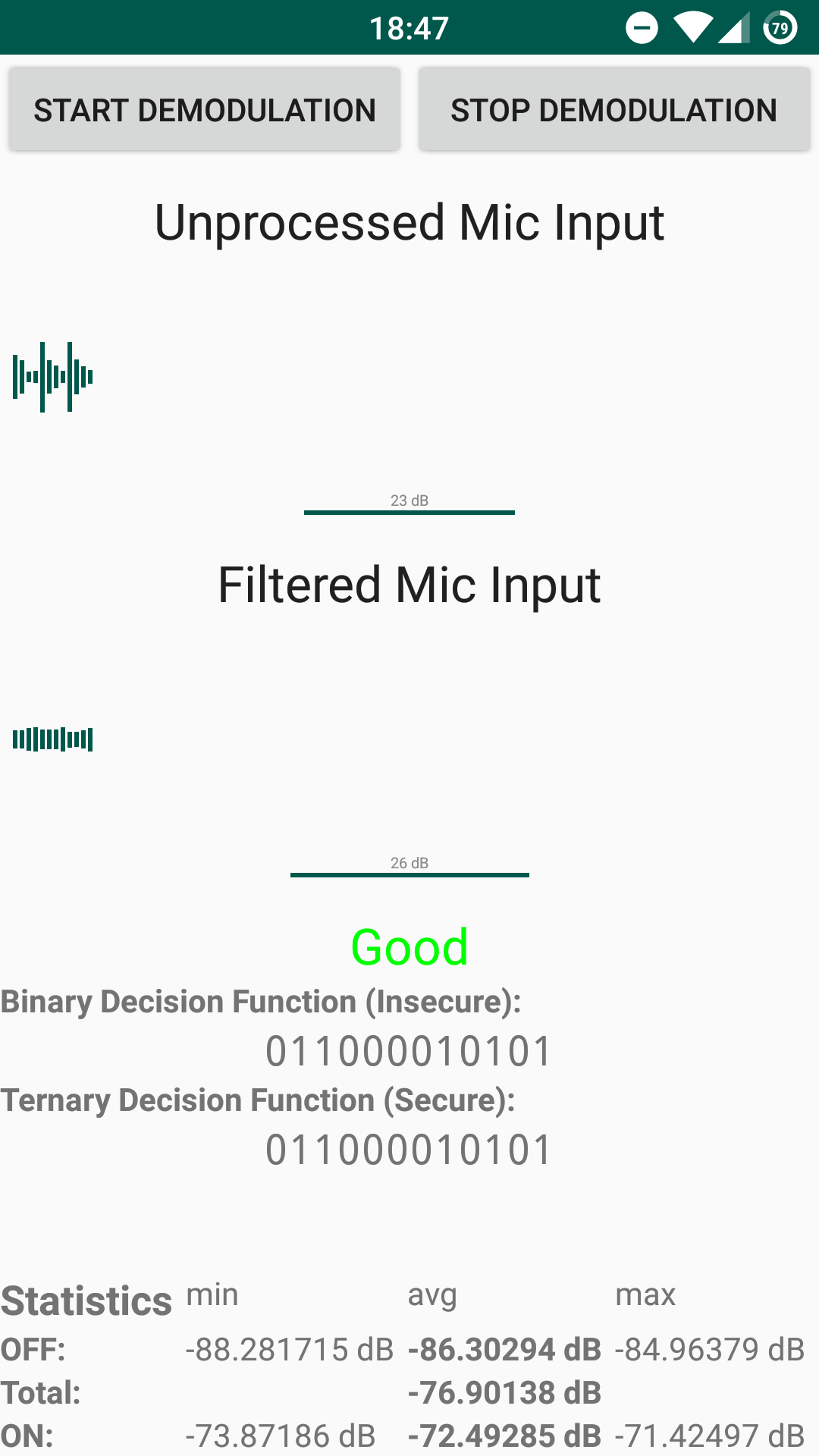}
    \Description{Android screenshot showing two buttons labelled ``Start
      demodulation'' and ``Stop demodulation'', and two plots labelled
      ``Unprocessed mic input'' and ``Filtered mic input''. The plots display
      waveforms. Below is a label indicating good signal quality. Below is a
      label showing ``Binary decision function (insecure)'' and the bits
      ``011010011010''. Below is a label showing ``Ternary decision function
      (secure)'' and the bits ``011010011010''. Below is an area displaying
      statistics of the received signal, consisting of a table showing minimum,
      average, and maximum powers of the off slots, on slots, and total signal.
      The average off slot power is ca. -86 dB and the average on slot power is
      ca. -72 dB.}
    \caption{Reception}
    \label{fig:android-rxfragment}
  \end{subfigure}
  \caption[Android Application]{Main views of the Android prototype.}
  \label{fig:android-screenshots}
\end{figure}

Our application consists of two components: (1) An Android library that handles
the modulation, transmission, 
demodulation, and reception of \ac{aic} signals, and (2) an Android module for the
user interface, which imports the library. We generate \ac{aic} signals by filling a
sample buffer with Gaussian distributed random numbers for each on slot and
applying a bandpass filter.
We transmit this \ac{aic} signal using the Android API
\texttt{android.audio.AudioTrack} in streaming mode. For reception we use the 
audio processing pipeline from \textit{TarsosDSP},
which is a Java framework for real-time audio analysis \cite{Six2015,Six2014}.
\autoref{fig:android-screenshots} shows the Android prototype's user interface. 

We tested our implementation on various Android smartphones: Huawei 
Nexus 6P (Android 8.1.0), LG G4 (Android 8.1.0), LG G5 (Android 7.1.2), LG Nexus
5 (Android 7.1), OnePlus 3T (Android 7.1.2), Samsung Galaxy S4 Mini (Android 9),
Samsung Galaxy S6 (Android 7.1.2), Xiaomi Redmi K20 Pro (Android 10). 

\section{Security Analysis}\label{sec:security-analysis}
In this section, we analyze if regular Integrity Codes and \acp{aic} satisfy our security goal of
providing \textit{message authentication}. A passive attack, where Mallory
only eavesdrops on the communication between $A$ and $B$, does not impact this
security goal. The adversary can perform
active attacks by sending signals,
which $B$ receives as $m'(t) = \sum_n m'_n(t)$  as part of $y(t)$ according to
\autoref{eq:received-signal}. $B$ requires
a high \ac{snr} to decode the signal:
\begin{align}\label{eq:snr}
  \text{SNR} = \frac{P_{\text{signal}}}{P_{\text{noise}}} = \frac{\int_0^T \left( x'(t) + m'(t) \right)^2\,\mathrm{d}t}{\int_0^T v(t)^2\,\mathrm{d}t}.
\end{align}
First, Mallory can try to disable communication between $A$ and $B$ via stateless
jamming, which reduces the SNR by increasing $P_{\text{noise}}$. 
Second, Mallory can perform \textit{signal cancellation}\footnote{Signal
  cancellation is also known as \textit{signal annihilation}.} attacks, by sending
signals that destructively interfere with the legitimate signal and, thus,
preventing $B$ from successfully decoding $A$'s signal. This attack reduces the
SNR by decreasing $P_{\text{signal}}$. A special case of this attack is
\textit{bit flipping}, where Mallory also sends her own message in addition to
signal cancellation.
Third, Mallory can perform \textit{overshadowing} attacks by sending her signals
with a power much higher than $A$, so that in the superposition at $B$ the
decoding process will be mostly influenced by Mallory.

We do not consider stateless jamming attacks, since protecting the availability of our
system is not our security goal. As our system does not rely on the
audibility of Mallory's signals, it is also not vulnerable to 
inaudible attacks~\cite{Roy2018, Zhang2017}. This leaves us with two distinct attack
vectors on the message's integrity: signal cancellation and overshadowing.

\subsection{Signal Cancellation Attacks}\label{sec:signal-cancellation}
\begin{figure}[!b]
  \centering
  \includegraphics[page=3,width=\columnwidth]{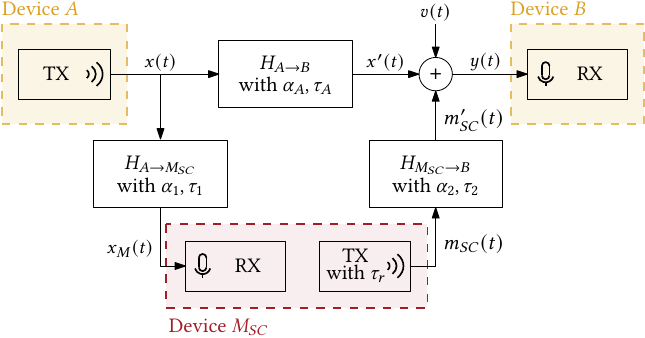}
  \Description{Block diagram showing the notation of the signals and channels
    between legitimate devices A and B and malicious device $M_{SC}$. }
  \caption[Block Diagram for Signal Cancellation]{Adversarial
    device $M_{SC}$ performing a signal cancellation attack.
  } 
  \label{fig:sig-cancellation-adversary-block}
\end{figure}

Integrity Codes rely on the assumption that signal cancellation is not
possible~\autocite{Capkun2008}, 
i.e., that it is impossible to perform a bit flip $1 \to 0$ in a signal modulated
using Integrity Codes. 
Under this assumption, any other modification $0 \to 1$ can be detected via
unidirectional error coding, which protects the integrity of 
the message. 
Signal cancellation attacks have
recently gained interest in the research
community~\autocite{Ghose2018,Moser2019,Poepper2011}. 
\autoref{fig:sig-cancellation-adversary-block} shows an exemplary signal
cancellation attack. Mallory uses device $M_{SC}$ (see adversary model in \autoref{fig:system-model}) to send a cancellation signal
$m_{SC}(t)$, which minimizes the power $P_{\text{signal}}$ received at $B$:
\begin{align}\label{eq:sc-power-signal}
  P_{\text{signal}} &= \frac{1}{T} \int_0^T \left( x'(t) + m'_{SC}(t) \right)^2 \, \text{d}t .
\end{align}
Destructive interference means that two waves with opposite polarity
superimpose~\cite[pp. 212-213]{Rossing2007}. For sound pressure waves, this is also known as
\textit{active noise cancellation (ANC)}.

For the purpose of blocking
communication by reducing the \ac{snr} at the receiver, signal cancellation is
more challenging to perform compared to jamming, since it requires the attacker
to both:
\begin{enumerate}
  \item predict her channel $ H_{M_{SC} \to B}$ to $B$, and 
  \item predict the
    signal $x'(t)$ from $A$ at $B$'s microphone.
\end{enumerate}
Mallory can then generate a signal $m_{SC}(t)$, which destructively interferes
and cancels $A$'s signal $x'(t)$.
Practical signal cancellation attacks have been demonstrated in lab
environments~\autocite{Moser2019,Poepper2011}. These attacks are challenging to
perform and require precise synchronization, especially when canceling
high-frequency signals. In the following security analysis, we assume a best
case scenario for the attacker, where she is able to completely predict all
channels. We use the approximate channel model from \autoref{sec:design} with
constant attenuation $\alpha$ and group delay $\tau$.

Mallory's goal is to construct a cancellation signal
\begin{align}
  m'_{SC}(t) = - x'(t - \tau_M) \label{eq:cancellation-signal}
\end{align}
with minimal \textit{cancellation delay} $\tau_M$, which requires predicting $x'(t)$.
Moser et al.
demonstrated a practical signal cancellation attack on predictable GPS
signals in a lab environment~\autocite{Moser2019}.
Mallory, however, cannot \textit{directly}
predict $x'(t)$ a priori without observing $A$'s signal $x(t)$, since
$x(t)$ is a stochastic signal.

We assume that Mallory can only \textit{indirectly} predict $x'(t)$ a posteriori using
her received version $x_M(t)$ of $x(t)$.
She therefore uses past values of $x_M(t)$
to predict future values of $x'(t)$.
\citeauthor{Poepper2011} demonstrated such a signal cancellation attack on QPSK signals in a static lab
environment~\autocite{Poepper2011}. They used two directional antennas as relays
(\textit{relaying attacker}) and relayed $A$'s signal $x_M(t)$ to $B$.

We now formalize this attack.
Mallory wants to relay the signal
\begin{align}
  x_M(t) = H_{A \to M_{SC}}\left\{ x(t) \right\} = \alpha_1 x(t - \tau_1).
\end{align}
The channel $H_{A \to M_{SC}}$ delays the signal $x(t)$ by $\tau_1$. When
Mallory relays this signal, the channel $H_{M_{SC} \to B}$ will delay it again
by $\tau_2$. She cannot invert these delays in real-time, since she does not have access to future
values of $x_M(t)$. She can only invert the attenuation of these channels. She
sends the signal 
\begin{align}
  m_{SC}(t) = - \frac{\alpha_A}{\alpha_1 \alpha_2} x_M(t - \tau_r),
\end{align}
where $\tau_r \geq \SI{0}{s}$ is an additional delay that she can freely adjust to
achieve better signal cancellation. $B$ then receives
\begin{align}
  m'_{SC}(t) &= H_{M_{SC} \to B}\left\{ m_{SC}(t) \right\}\\
             &= - \frac{\alpha_A}{\alpha_1} x_M(t - \tau_2 - \tau_r)\nonumber\\
             &= - \alpha_A x(t - \tau_1 - \tau_2 - \tau_r).\nonumber
\end{align}
We can rewrite this using \autoref{eq:x-prime} as
\begin{align}
  m'_{SC}(t) = - x'(t + \tau_A - \tau_1 - \tau_2 - \tau_r).
\end{align}
For indirect prediction of $x'(t)$, the cancellation delay (according to \autoref{eq:cancellation-signal}) therefore is
\begin{align}
 \tau_M = \tau_1 + \tau_2 + \tau_r - \tau_A. \label{eq:tau-m-relay}
\end{align}

We now analyze whether a relay attack is possible against \ac{aic} signals.
Since we require Mallory to
operate outside the safe area $S$, we can give a lower bound for the
cancellation delay due to the propagation delay:
\begin{align}
  \tau_M = \tau_r + \frac{\overline{A M_{SC} B} - \overline{AB}}{c_s} \geq \frac{2r - \overline{AB}}{c_s} \label{eq:tau-m-lower-bound}
\end{align}
where $c_s$ is the speed of sound in the transmission medium.
The
cancellation delay $\tau_M$ increases for larger safe area radii $r$ and for
smaller $\overline{AB}$. For example, Mallory can achieve $\tau_M > \SI{1}{ms}$ for realistic parameters $r >
\SI{40}{cm} > \overline{AB}$. The sound wave's speed
is an inherent physical-layer limitation that passively assists us with authentication. A
related strategy is distance bounding~\autocite{Brands1994}, which actively measures the propagation
delay and therefore requires more sophisticated implementations.

We measure the effect of this attack using the resulting power
\begin{align}
  P_{signal} &= \frac{1}{T} \int_0^T \left( x'(t) - x'(t - \tau_M) \right)^2 \, \text{d}t\\
             &= 2P_{x'} - 2\frac{1}{T} \int_0^T x'(t)x'(t - \tau_M) \, \text{d}t .\nonumber
\end{align}
Note that without any cancellation delay ($\tau_M = 0$), Mallory would be able
to completely cancel the signal $x'(t)$. Otherwise, Mallory has to \textit{maximize} the
subtrahend, which contains the autocorrelation of the signal $x'(t)$:
\begin{align}
  R_{x'x'}(\tau_M) = \int x'(t)x'(t - \tau_M) \, \text{d}t .
\end{align}
This is related to the autocorrelation of the original signal $x(t)$:
\begin{align}
  R_{x'x'}(\tau_M) = \alpha_A^2 \int x(t - \tau_A)x(t - \tau_A - \tau_M) \, \text{d}t = \alpha_A^2 R_{xx}(\tau_M).\label{eq:autocorrelation-relation}
\end{align}

To defend against relay attacks, we therefore have to construct \ac{aic} signals with
\textit{low autocorrelation} $R_{xx}(\tau_M) \approx 0$ for $\tau_M > \SI{1}{ms}$.
The on slots in \ac{aic} signals consist of a \textit{stochastic ``carrier'' signal}
$w(t)$, which we generate by sampling from the stochastic process $\{W(t)\}$.
We implement \acp{aic} using Gaussian distributed signals $\{W_{WGN}(t)\}$ for the
on slots, which is optimal because white gaussian noise has minimal
autocorrelation~\autocite[p.~189]{Proakis2002}.

To simplify the implementation, most publications on Integrity Codes use an
existing modulation scheme with random symbols as the on slots:
FSK~\autocite{Capkun2008}, QPSK~\autocite{Hou2015}, or OFDM in combination with
QAM~\autocite{Capkun2008,Gollakota2011,Pan2017}. This simplifies 
the implementation because parts of an existing physical-layer pipeline, such as
a Wi-Fi chip or an SDR reference implementation, can be reused. It is a
security tradeoff, though, since these modulation schemes usually have high
autocorrelation, depending on the slot size $T_S$, which in turn is prone to
signal cancellation.

\begin{figure}[!b]
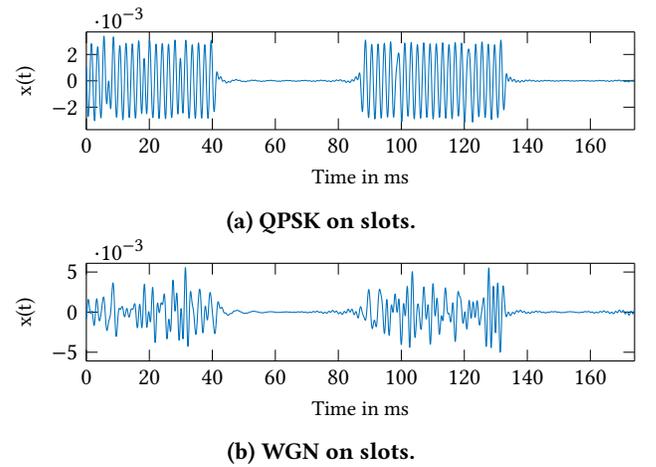

  \begin{subfigure}{\columnwidth}
    \centering
    \setlength\figureheight{1.3cm}
    \setlength\figurewidth{0.86\columnwidth}
    \input{gfx/aic_qpsk.tikz}
    \Description{Plot with time on the horizontal axis and signal intensity on
      the vertical axis. From 0 ms to 40 ms the signal is oscillating quite
      regularly. From 40 ms to 85 ms the signal is zero. From 85 ms to 130 ms
      the signal is oscillating quite regularly again. From 130 ms to 170 ms the
    signal is zero.}
    \caption{QPSK on slots.}
    \label{fig:aic-qpsk}
  \end{subfigure}
  \begin{subfigure}{\columnwidth}
    \centering
    \setlength\figureheight{1.3cm}
    \setlength\figurewidth{0.86\columnwidth}
    \input{gfx/aic_gaussian.tikz}
    \Description{Plot with time on the horizontal axis and signal intensity on
      the vertical axis. From 0 ms to 40 ms the signal varies randomly. From 40
      ms to 85 ms the signal is zero. From 85 ms to 130 ms the signal varies
      randomly again. From 130 ms to 170 ms the signal is zero.}
    \caption{WGN on slots.}
    \label{fig:aic-gaussian}
  \end{subfigure}
  \caption{Four slots of an \ac{aic} signal using different on slot implementations.}
  \label{fig:aic-qpsk-gaussian}
\end{figure}

\subsection{Evaluation of Signal Cancellation Attacks}
We compare two choices of $\{W(t)\}$ using \MATLAB{} simulations:
\begin{enumerate}
\item On slots containing \textit{QPSK signals}, which have high
  autocorrelation. Most other implementations of
  Integrity Codes use an existing modulation scheme such as QPSK.
\item On slots containing \textit{Gaussian distributed signals}, which have
  low autocorrelation. This corresponds to our implementation of \acp{aic}.
\end{enumerate}

For both cases, we
measure the autocorrelation coefficient and the attenuation that a signal
cancellation attacker achieves for different cancellation delays $\tau_M$. Our
evaluation applies to different device distances $\overline{AB}$ and attacker
locations according to \autoref{eq:tau-m-lower-bound}.
Mallory
aims to achieve high attenuation to cancel Alice's signal.
We do not vary the SNR, because we assume a best-case scenario for Mallory where
she is able to match Alice's SNR. We use the frequency band $[\SI{200}{Hz},
\SI{800}{Hz}]$ to better visualize the security impact. For higher frequencies,
signal cancellation is even more challenging due to stricter timing constraints.

\subsubsection{QPSK On Slots}

\autoref{fig:aic-qpsk} shows an example of an \ac{aic} signal $x(t)$ using the random
process $\{W_{QPSK}(t)\}$, which is a non-stationary random process containing
random QPSK symbols (drawn independently with uniform probability). For this
example, each QPSK symbol has duration $T_Q = \frac{T_S}{4}$, such that each
slot contains four QPSK symbols.
This use of
\textit{``minislots''} increases 
the security~\autocite{Capkun2008}, by reducing the autocorrelation. We use a
gross bit rate $R_g \approx \SI{21.8}{bps}$ and a carrier
frequency $f_c =  \SI{500}{Hz}$. 

Even though the content of each on slot is ``random'' in the sense
that it consists of four QPSK symbols, where each symbol was independently drawn
from one of four possible QPSK symbols, it is still deterministic during each of
these QPSK symbols. The underlying period $f_c$ of the deterministic carrier signal
can be clearly seen. Each QPSK symbol only carries two bits of information.

\begin{figure}[!b]
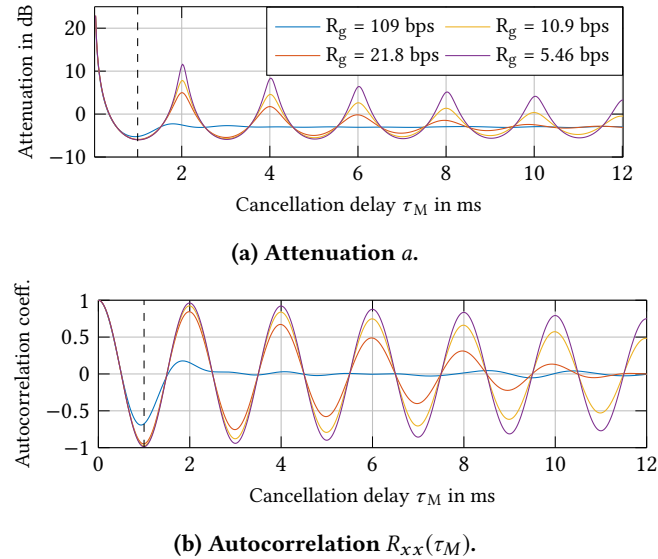

  \begin{subfigure}{\columnwidth}
    \centering
    \setlength\figureheight{2cm}
    \setlength\figurewidth{0.86\textwidth}
    \input{gfx/aic_qpsk_attacker_attenuations.tikz}
    \Description{Plot showing the cancellation delay from 0 ms to 12 ms on the
      horizontal axis and the attenuation in dB on the vertical axis. The plot
      contains 4 lines corresponding to different gross bit rates. The line for
      a low gross bit rate of 5.46 bps has multiple peaks at 0 ms, 2 ms, 4 ms, 6
      ms, 8 ms, 10 ms and 12 ms. The attenuation reaches 5 dB to 10 dB at these
      peaks. Between the peaks, the attenuation becomes negative. The line
      corresponding to a high gross bit rate of 109 bps stays in the negative
      area after less then 1 ms. A vertical dashed line at 1 ms corresponds to
      realistic safe area parameters.}
    \caption{Attenuation $a$.}
    \label{fig:aic-qpsk-attenuation}
  \end{subfigure}
  \begin{subfigure}{\columnwidth}
    \centering
    \setlength\figureheight{2cm}
    \setlength\figurewidth{0.86\textwidth}
    \input{gfx/aic_qpsk_autocorrelation.tikz}
    \Description{Plot showing the cancellation delay from 0 ms to 12 ms on the
      horizontal axis and the autocorrelation coefficient on the vertical axis. The plot
      contains 4 lines corresponding to different gross bit rates. The line for
      a low gross bit rate of 5.46 bps looks like a slightly damped cosine wave
      and has multiple peaks at 0 ms, 2 ms, 4 ms, 6
      ms, 8 ms, 10 ms and 12 ms. The autocorrelation coefficient reaches 0.7 to 1 at these
      peaks. The line
      corresponding to a high gross bit rate of 109 bps oscillates at first but
      settles at around 0 after 1 ms to 2 ms. A vertical dashed line at 1 ms corresponds to
      realistic safe area parameters.}
    \caption{Autocorrelation $R_{xx}(\tau_M)$.}
    \label{fig:aic-qpsk-autocorrelation}
  \end{subfigure}
  \caption[\ac{aic} Attenuation, Autocorrelation Using QPSK]{Attenuation and
    Autocorrelation of an \ac{aic} signal using $\{W_{QPSK}(t)\}$.}
  \label{fig:aic-qpsk-signal-cancellation}
\end{figure}

\autoref{fig:aic-qpsk-attenuation} shows the attenuation that Mallory achieves
for different time delays using a relay attack.
Destructive interference occurs at multiples of the carrier period
$\frac{1}{f_c} = \SI{2}{ms}$, which is possible even for realistic values of $\tau_M
> \SI{1}{ms}$ (dashed line).
For most time delays and for high bit rates, however, the signals interfere constructively and the
attenuation is $< \SI{0}{dB}$.  Mallory therefore has to precisely control her 
additional delay $\tau_r$ in \autoref{eq:tau-m-relay}.

\autoref{fig:aic-qpsk-autocorrelation} shows the autocorrelation
$R_{xx}(\tau_M)$ of an \ac{aic} signal using $\{W_{QPSK}(t)\}$. The autocorrelation has the same peaks as the attenuation in
\autoref{fig:aic-qpsk-attenuation}, at multiples of the carrier period
$\frac{1}{f_c} = \SI{2}{ms}$. This is consistent with our argument that signal
cancellation requires high autocorrelation.

\subsubsection{Gaussian On Slots}
\autoref{fig:aic-gaussian} shows an example of an \ac{aic} signal $x(t)$ using a white
Gaussian noise process $\{W_{WGN}(t)\}$. Compared to the QPSK-shaped
on slots, the Gaussian-shaped on slots carry more 
information and are therefore harder to predict. We cannot determine any obvious
patterns when looking at the plot.

\begin{figure}[!b]
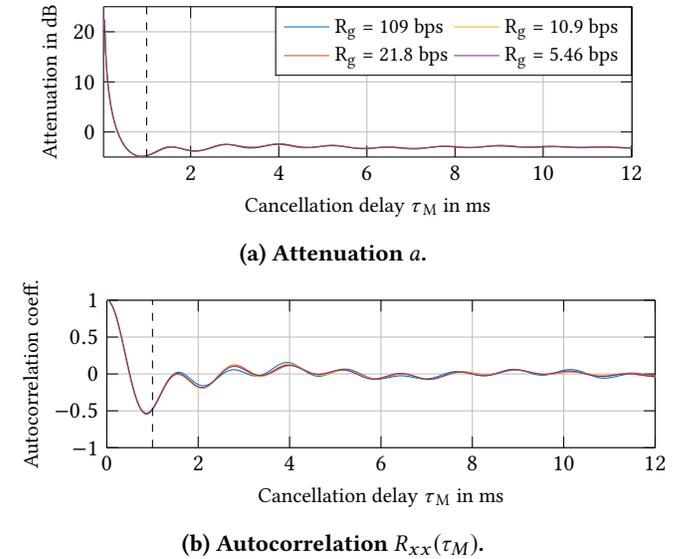

  \begin{subfigure}{\columnwidth}
    \centering
    \setlength\figureheight{2cm}
    \setlength\figurewidth{0.86\textwidth}
    \input{gfx/aic_gaussian_attacker_attenuations.tikz}
    \Description{Plot showing the cancellation delay from 0 ms to 12 ms on the
      horizontal axis and the attenuation in dB on the vertical axis. The plot
      contains 4 lines corresponding to different gross bit rates. All lines
      look nearly identical. The attenuation becomes negative after less than 1
      ms. A vertical dashed line at 1 ms corresponds to
      realistic safe area parameters.}
    \caption{Attenuation $a$.}
    \label{fig:aic-gaussian-attenuation}
  \end{subfigure}
  \begin{subfigure}{\columnwidth}
    \centering
    \setlength\figureheight{2cm}
    \setlength\figurewidth{0.86\textwidth}
    \input{gfx/aic_gaussian_autocorrelation.tikz}
    \Description{Plot showing the cancellation delay from 0 ms to 12 ms on the
      horizontal axis and the autocorrelation coefficient on the vertical axis. The plot
      contains 4 lines corresponding to different gross bit rates. All lines
      look nearly identical. The autocorrelation fluctuates close at zero after
      1 ms to 2 ms. A vertical dashed line at 1 ms corresponds to
      realistic safe area parameters.}
    \caption{Autocorrelation $R_{xx}(\tau_M)$.}
    \label{fig:aic-gaussian-autocorrelation}
  \end{subfigure}
  \caption[\ac{aic} Attenuation, Autocorrelation Using WGN]{Attenuation and
    Autocorrelation of an \ac{aic} signal using $\{W_{WGN}(t)\}$.} 
  \label{fig:aic-gaussian-signal-cancellation}
\end{figure}

\autoref{fig:aic-gaussian-attenuation} shows the attenuation that Mallory achieves
for different time delays using a relay attack.
The attenuation is positive only for
$\tau_M \approx 0$, where Mallory can successfully cancel the \ac{aic} signal.
\autoref{fig:aic-gaussian-autocorrelation} shows the corresponding
autocorrelation $R_{xx}(\tau_M)$ of the \ac{aic} signal. The autocorrelation coefficient is
approximately zero for $\tau_M > \SI{1.5}{ms}$.
When using a higher frequency band, such as $[\SI{16}{kHz}, \SI{20}{kHz}]$,
Mallory needs to achieve even lower cancellation delays $\tau_M \ll \SI{1}{ms}$.
As this is not possible for realistic safe areas, Mallory's cancellation signal
actually increases the received power.

\subsection{Overshadowing Attacks}\label{sec:overshadowing}
In overshadowing attacks, Mallory attempts to send her own \ac{aic} signal $m_O(t)$
(see \autoref{fig:system-model}) with much
greater power than the legitimate signal $x(t)$, so that her signal $m_O(t)$ determines
the data that $B$ decodes. In contrast to signal cancellation attacks, Mallory
does not necessarily need to receive the legitimate signal. She could use it,
however, to obtain timing information and synchronize with the legitimate
signal.

We 
defend against overshadowing attacks by adjusting the receiver's detection step.
The receiver measures the power 
$\hat{c}_p[n]$ of every slot and applies a decision function $D: \mathbb{R}
\times \mathbb{R} \mapsto S$ on every Manchester pair to determine the bit
$\hat{d}[n]$ that this pair encodes:
\begin{align}
  \hat{d}[n] = D\left( \hat{c}_p[2n], \hat{c}_p[2n + 1] \right).
\end{align}
Overshadowing attacks result in high slot powers in both slots, which our
decision function $D_{\text{ternary}}$ (see \autoref{eq:d-ternary}) detects due
to the threshold $P_{\text{th}}$. The threshold should be as low as
possible but definitely lower than
Alice's average signal power. If an overshadowing attacker attempts a bit flip, her
off slot overlaps with the legitimate on slot, which has higher power than
$P_{\text{th}}$. For conventional modulation schemes such as \ac{ask}, however,
existing decoders do not consider an attacker at the physical layer 
and instead aim at maximizing robustness to increase throughput. These binary
decision functions $D_{\text{binary}}$ decode the Manchester encoded slots based on a relative
comparison between the slot powers, which is vulnerable to overshadowing
attacks. 

\begin{table}[!t]
  \centering
  \caption[Comparison of Decision Functions]{Comparison of $D_{\text{binary}}$ and $D_{\text{ternary}}$ in the presence of an
    overshadowing attacker transmitting $d_M$ instead of the legitimate $d$. The
    highlighted rows indicate attempted bit flips.}
  \label{tab:detection-overshadowing-attacker}
  \begin{tabular}{@{}>{\columncolor{white}[0pt][\tabcolsep]}cccccc>{\columncolor{white}[0pt][\tabcolsep]}l@{}}
    \toprule
    ~$d$ & $d_M$ & $p_1 / \si{dBFS}$ & $p_2 / \si{dBFS}$ & $D_{\text{binary}}(p_1, p_2)$ & $D_{\text{ternary}}(p_1, p_2)$\\
    \midrule
    ~0 & - & -90 & -70 & 0 & 0 \\
    ~0 & 0 & -90 & -59.6 & 0 & 0 \\
    \rowcolor{red!20}~0 & 1 & -60 & -70 & 1 & $\epsilon$ \\
    \midrule
    ~1 & - & -70 & -90 & 1 & 1 \\
    ~1 & 1 & -59.6 & -90 & 1 & 1 \\
    \rowcolor{red!20}~1 & 0 & -70 & -60 & 0 & $\epsilon$ \\
    \bottomrule
  \end{tabular}
\end{table}
\subsection{Evaluation of Overshadowing Attacks}
We consider an overshadowing attacker in an exemplary scenario with a noise
floor of \SI{-90}{dBFS}, an SNR of \SI{20}{dB} for the legitimate sender, a
higher SNR of \SI{30}{dB} for the overshadowing attacker, and a threshold of
\SI{-80}{dBFS} at the receiver.
\autoref{tab:detection-overshadowing-attacker} lists all
combinations of legitimate message bit $d$ and adversarial message bit $d_M$.
The receiver can detect bit flips (highlighted rows) using the secure
decision function $D_{\text{ternary}}$, while $D_{\text{binary}}$ is vulnerable.

\section{Robustness Evaluation}\label{sec:evaluation}
\acreset{snr}
\acreset{ber}
In the last section, we considered an active adversary attacking our system. We
now evaluate the robustness of \acp{aic} when there is no active attacker. In this
case, we have to deal with background noise on the acoustic channel.

\subsection{Methodology}
We use our \MATLAB{} implementation to simulate AIC signals and background noise. We vary the
\ac{snr} at the receiver (see \autoref{eq:snr}), the gross bit rate $R_g = \frac{1}{T_s}$, the detection threshold
$\text{SNR}_{\text{th}} = \frac{P_{\text{th}}}{P_{\text{Noise}}}$,
and the bandwidth of our system. Our evaluation results do not depend on the
transmit power or the device distance $\overline{AB}$, as these parameters
both influence the resulting SNR.
We measure the
resulting \ac{ber}.

All simulations use a sample rate of $f_s = \SI{44.1}{kHz}$. We
simulate additive white Gaussian noise with a noise power of
\SI{-87}{dBFS} in 
the frequency band. We transmit \SI{128}{bits} of random data $d$. We repeat all
experiments 200 times and give the average result. 

\subsection{Inter-Symbol Interference}\label{sec:isi}

The bandwidth influences how fast the transitions between on slots and off slots
can happen. For low bandwidths, this transition takes longer and the on
slots' energy spreads to the off slots, which is called \textit{inter-symbol
  interference (ISI)}. 
We observe that both very low and very high SNRs suffer from low bandwidth. This occurs for different reasons:
\begin{itemize}
  \item 
    For \textit{low signal powers} below the detection
    threshold, the limited bandwidth further reduces the 
    power in the on slots, leading to detection difficulties at the receiver.
  \item
    For \textit{high signal powers} well above the detection threshold, the
    limited bandwidth leads to increased 
    power in 
    the off slots due to ISI. On slots and off slots are then 
    above the detection threshold, which is rejected by the decision function.
\end{itemize}

\subsection{Results}\label{sec:evaluation-results}
We achieve low BERs
below \SI{1}{\percent} for $R_g < \SI{450}{bps}$ and $\text{SNR} >
\text{SNR}_{\text{th}} = \SI{11}{dB}$ using bandwidths larger than
\SI{4}{kHz}. For a net bit rate of \SI{100}{bps}, e.g., we achieve a BER below
\SI{0.1}{\percent} at an SNR of \SI{14}{dB}.
This corresponds to a transmission time of
approximately \SI{1.2}{s} for a \SI{128}{bit} hash value.
The BER increases for higher bit rates or lower bandwidths due to
ISI. Using the insecure decision function
$D_{\text{binary}}$ improves robustness at the cost of being vulnerable to overshadowing attacks
(see \autoref{sec:overshadowing}).

\begin{figure}
    \centering
    \setlength\figureheight{4cm}
    \setlength\figurewidth{0.86\columnwidth}
%
%
\definecolor{mycolor1}{rgb}{0.00000,0.44700,0.74100}%
\definecolor{mycolor2}{rgb}{0.85000,0.32500,0.09800}%
\definecolor{mycolor3}{rgb}{0.92900,0.69400,0.12500}%
\definecolor{mycolor4}{rgb}{0.49400,0.18400,0.55600}%
\definecolor{mycolor5}{rgb}{0.46600,0.67400,0.18800}%
\definecolor{mycolor6}{rgb}{0.30100,0.74500,0.93300}%
\begin{tikzpicture}

\begin{axis}[%
width=0.951\figurewidth,
height=\figureheight,
at={(0\figurewidth,0\figureheight)},
scale only axis,
separate axis lines,
every outer x axis line/.append style={black},
every x tick label/.append style={font=\color{black}},
every x tick/.append style={black},
xmin=0.000000,
xmax=25.000000,
xtick={0.000000,5.000000,10.000000,15.000000,20.000000,25.000000},
xlabel={$\text{SNR}_{\text{th}}\text{ in dB}$},
every outer y axis line/.append style={black},
every y tick label/.append style={font=\color{black}},
every y tick/.append style={black},
ymin=0.000000,
ymax=100.000000,
ytick={0.000000,10.000000,20.000000,30.000000,40.000000,50.000000,60.000000,70.000000,80.000000,90.000000,100.000000},
ylabel={$\text{BER}_\text{A}\text{ in \%}$},
axis background/.style={fill=white},
legend style={legend cell align=left, align=left, draw=white!15!black},
clip mode=individual,transpose legend,legend columns=2,legend style={at={(0.5,1.06)},anchor=south,draw=black,fill=white, font=\small},tick label style={font=\tiny},label style={font=\small}, title style={font=\small\bfseries},
]
\addplot [color=mycolor1, mark=x, mark options={solid, mycolor1}]
  table[row sep=crcr]{%
0.000000	74.427083\\
1.000000	35.026042\\
2.000000	10.026042\\
3.000000	0.442708\\
4.000000	0.000000\\
5.000000	0.000000\\
6.000000	0.000000\\
7.000000	0.000000\\
8.000000	0.000000\\
9.000000	0.026042\\
10.000000	0.312500\\
11.000000	2.760417\\
12.000000	13.151042\\
13.000000	44.427083\\
14.000000	84.218750\\
15.000000	98.046875\\
16.000000	100.000000\\
17.000000	100.000000\\
18.000000	100.000000\\
19.000000	100.000000\\
20.000000	100.000000\\
21.000000	100.000000\\
22.000000	100.000000\\
23.000000	100.000000\\
24.000000	100.000000\\
25.000000	100.000000\\
};
\addlegendentry{$\text{SNR}_\text{A}\text{ = 10 dB}$}

\addplot [color=mycolor1, dashed, forget plot]
  table[row sep=crcr]{%
13.000000	0.000000\\
13.000000	100.000000\\
};
\addplot [color=mycolor2, mark=o, mark options={solid, mycolor2}]
  table[row sep=crcr]{%
0.000000	79.166667\\
1.000000	42.057292\\
2.000000	13.203125\\
3.000000	1.223958\\
4.000000	0.000000\\
5.000000	0.026042\\
6.000000	0.000000\\
7.000000	0.000000\\
8.000000	0.000000\\
9.000000	0.000000\\
10.000000	0.000000\\
11.000000	0.026042\\
12.000000	0.390625\\
13.000000	3.515625\\
14.000000	19.479167\\
15.000000	54.921875\\
16.000000	92.526042\\
17.000000	98.828125\\
18.000000	100.000000\\
19.000000	100.000000\\
20.000000	100.000000\\
21.000000	100.000000\\
22.000000	100.000000\\
23.000000	100.000000\\
24.000000	100.000000\\
25.000000	100.000000\\
};
\addlegendentry{$\text{SNR}_\text{A}\text{ = 12 dB}$}

\addplot [color=mycolor2, dashed, forget plot]
  table[row sep=crcr]{%
15.000000	0.000000\\
15.000000	100.000000\\
};
\addplot [color=mycolor3, mark=square, mark options={solid, mycolor3}]
  table[row sep=crcr]{%
0.000000	84.270833\\
1.000000	53.906250\\
2.000000	21.328125\\
3.000000	3.619792\\
4.000000	0.234375\\
5.000000	0.000000\\
6.000000	0.000000\\
7.000000	0.000000\\
8.000000	0.000000\\
9.000000	0.000000\\
10.000000	0.000000\\
11.000000	0.000000\\
12.000000	0.000000\\
13.000000	0.078125\\
14.000000	0.442708\\
15.000000	2.942708\\
16.000000	21.015625\\
17.000000	57.526042\\
18.000000	93.411458\\
19.000000	99.505208\\
20.000000	100.000000\\
21.000000	100.000000\\
22.000000	100.000000\\
23.000000	100.000000\\
24.000000	100.000000\\
25.000000	100.000000\\
};
\addlegendentry{$\text{SNR}_\text{A}\text{ = 14 dB}$}

\addplot [color=mycolor3, dashed, forget plot]
  table[row sep=crcr]{%
17.000000	0.000000\\
17.000000	100.000000\\
};
\addplot [color=mycolor4, mark=diamond, mark options={solid, mycolor4}]
  table[row sep=crcr]{%
0.000000	90.494792\\
1.000000	66.770833\\
2.000000	31.432292\\
3.000000	9.869792\\
4.000000	2.213542\\
5.000000	0.156250\\
6.000000	0.000000\\
7.000000	0.000000\\
8.000000	0.000000\\
9.000000	0.000000\\
10.000000	0.000000\\
11.000000	0.000000\\
12.000000	0.000000\\
13.000000	0.026042\\
14.000000	0.026042\\
15.000000	0.000000\\
16.000000	0.677083\\
17.000000	3.671875\\
18.000000	22.343750\\
19.000000	66.953125\\
20.000000	91.380208\\
21.000000	99.166667\\
22.000000	100.000000\\
23.000000	100.000000\\
24.000000	100.000000\\
25.000000	100.000000\\
};
\addlegendentry{$\text{SNR}_\text{A}\text{ = 16 dB}$}

\addplot [color=mycolor4, dashed, forget plot]
  table[row sep=crcr]{%
19.000000	0.000000\\
19.000000	100.000000\\
};
\addplot [color=mycolor5, mark=asterisk, mark options={solid, mycolor5}]
  table[row sep=crcr]{%
0.000000	95.104167\\
1.000000	78.645833\\
2.000000	50.755208\\
3.000000	20.833333\\
4.000000	7.343750\\
5.000000	1.640625\\
6.000000	0.234375\\
7.000000	0.000000\\
8.000000	0.000000\\
9.000000	0.000000\\
10.000000	0.000000\\
11.000000	0.000000\\
12.000000	0.000000\\
13.000000	0.000000\\
14.000000	0.000000\\
15.000000	0.000000\\
16.000000	0.000000\\
17.000000	0.052083\\
18.000000	0.494792\\
19.000000	4.140625\\
20.000000	18.932292\\
21.000000	57.760417\\
22.000000	88.593750\\
23.000000	99.322917\\
24.000000	99.973958\\
25.000000	100.000000\\
};
\addlegendentry{$\text{SNR}_\text{A}\text{ = 18 dB}$}

\addplot [color=mycolor5, dashed, forget plot]
  table[row sep=crcr]{%
21.000000	0.000000\\
21.000000	100.000000\\
};
\addplot [color=mycolor6, mark=triangle, mark options={solid, mycolor6}]
  table[row sep=crcr]{%
0.000000	97.239583\\
1.000000	86.354167\\
2.000000	67.421875\\
3.000000	43.307292\\
4.000000	20.755208\\
5.000000	7.031250\\
6.000000	2.187500\\
7.000000	0.364583\\
8.000000	0.130208\\
9.000000	0.000000\\
10.000000	0.000000\\
11.000000	0.000000\\
12.000000	0.000000\\
13.000000	0.000000\\
14.000000	0.000000\\
15.000000	0.000000\\
16.000000	0.000000\\
17.000000	0.000000\\
18.000000	0.000000\\
19.000000	0.000000\\
20.000000	0.338542\\
21.000000	6.901042\\
22.000000	21.822917\\
23.000000	55.807292\\
24.000000	93.932292\\
25.000000	99.453125\\
};
\addlegendentry{$\text{SNR}_\text{A}\text{ = 20 dB}$}

\addplot [color=mycolor6, dashed, forget plot]
  table[row sep=crcr]{%
23.000000	0.000000\\
23.000000	100.000000\\
};
\end{axis}

\begin{axis}[%
width=1.227\figurewidth,
height=1.227\figureheight,
at={(-0.16\figurewidth,-0.135\figureheight)},
scale only axis,
xmin=0.000000,
xmax=1.000000,
ymin=0.000000,
ymax=1.000000,
axis line style={draw=none},
ticks=none,
axis x line*=bottom,
axis y line*=left,
clip mode=individual,transpose legend,legend columns=2,legend style={at={(0.5,1.06)},anchor=south,draw=black,fill=white, font=\small},tick label style={font=\tiny},label style={font=\small}, title style={font=\small\bfseries},
]
\end{axis}
\end{tikzpicture}%
    \Description{This plot shows the detection threshold between 0 dB and 25 dB
      on the horizontal axis and the bit error rate betwen 0 and 1 on the
      vertical axis. The plot contains 6 different lines corresponding to
      different signal SNRs from 10 dB to 20 dB.}
    \caption{BER for different
      detection thresholds $\text{SNR}_{\text{th}}$ and $\text{SNR}$s.}
  \label{fig:evaluation-robustness-ber-threshold}
\end{figure}
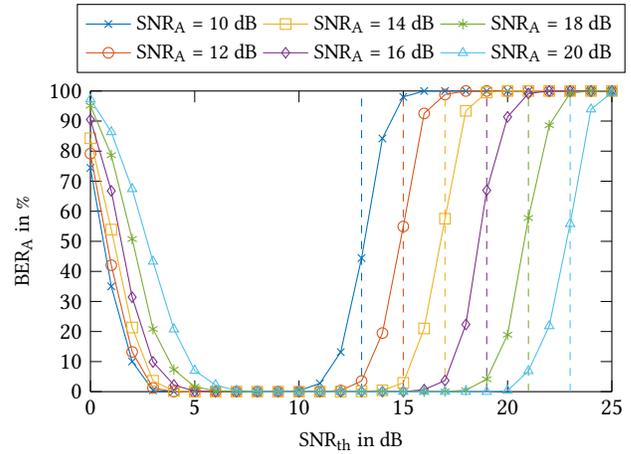

The detection threshold $\text{SNR}_{\text{th}}$ influences the robustness of our
pairing scheme using the ternary decision function $D_{\text{ternary}}$. In
\autoref{fig:evaluation-robustness-ber-threshold}, we plot the BER for different
SNRs and detection thresholds, using the gross bit rate 
$R_g = \SI{220}{bps}$.
For a fixed SNR, the BER first decreases to approximately \SI{0}{\percent} and then increases again
to \SI{100}{\percent}: 
\begin{itemize}
\item 
  For $\text{SNR}_{\text{th}} \approx \SI{0}{dB}$, ISI leads to bit
  errors because the on slots' energy spreads to the 
  off slots, which both surpass the detection threshold. This is why high SNRs,
  where the on slots contain more energy, perform worse in the 
  left half of the plot. 
\item
  The increase in the right half of the plot is due to the detection threshold
  surpassing the power of the on slots. This occurs at ca.
  $\text{SNR}_{\text{th}} \approx \text{SNR} + \SI{3}{dB}$ (dashed
  lines), due to the on slots having approximately twice the power than the
  average signal power SNR. 
\end{itemize}

The ISI in the plot's left half decreases for lower bit rates. 
Our results indicate that there is less tolerance on the
detection threshold for high bit rates and varying SNR values. For low bit
rates, the range of detection thresholds that allow error-free transmission
increases.

\section{Conclusion}\label{sec:conclusion}
\acreset{aic}
\acreset{snr}
\ac{sdp} relies on an OOB channel to authenticate devices. We designed an \ac{sdp} scheme
using short-range acoustic communication to transmit key material. We proposed
\emph{\acp{aic}} to achieve message authentication on the
acoustic physical layer. To the best
of our knowledge, Integrity Codes have not been used to secure acoustic
communication before.

Integrity Codes can be vulnerable to signal cancellation
attacks if the transmitted on slots are not sufficiently random.
Our security
analysis shows that we can defend against \emph{signal cancellation attacks} by 
designing signals with low autocorrelation, e.g., Gaussian distributed signals.
We introduced a set of realistic operation parameters that mitigate signal
cancellation attacks via additional propagation delays:
Compared to conventional modulation schemes that do not consider an attacker at the
physical layer, our system can also detect \emph{overshadowing attacks} by using a 
threshold in the receiver's decision function. The attacker cannot
impersonate the legitimate sender. 

The robustness of \acp{aic} depends on the channel conditions and the
desired bit rate.
Our evaluation demonstrated that lower
bit rates, higher \acp{snr}, and higher bandwidths improve the bit error rate.
Finally, we implemented a \emph{proof-of-concept} for Android devices to demonstrate
practical pairing between different smartphone models.

\begin{acks}
This work has been co-funded by the LOEWE initiative (Hessen State Ministry
for Higher Education, Research and the Arts, Germany) within the emergenCITY
centre and by the German Federal Ministry of Education and Research and the
Hessen State Ministry for Higher Education, Research and the Arts within their
joint support of the National Research Center for Applied Cybersecurity
ATHENE. 
\end{acks}

\printbibliography

\end{document}